\documentclass{aa501}
\usepackage{graphicx}
\newcommand{\kmprs}  {\mbox{\rm \, km\,\,s$^{-1}$}}
\newcommand{\mA} {\mbox {m\AA}}
\newcommand{\feh} {\mbox{\rm [Fe/H]}}
\newcommand{\nlia} {\log \varepsilon}
\newcommand{\nli} {\log \varepsilon ({\rm Li})}

\newcommand{\teff}  {\mbox{$T_{\rm eff}$}}
\newcommand{\logteff} {\mbox{${\rm log} T_{\rm eff}$}}
\newcommand{\logg}  {\mbox{{\rm log}$g$}}
\newcommand{\ffe}  {\mbox{${\rm [\frac{Fe}{H}]}$}}
\newcommand{\Lione} {\ion{Li}{i}}
 
\newcommand{\Feone} {\ion{Fe}{i}}
\newcommand{\Fetwo} {\ion{Fe}{ii}}

\newcommand{\Mv} {\mbox{$M_V$}}

\newcommand{\Lisix} {\element[][6]{Li}}
\newcommand{\Liseven} {\element[][7]{Li}}
\newcommand{\sixseven} {\element[][6]{Li}/\element[][7]{Li}}
\begin{document}

\title{Lithium abundances for 185 main-sequence stars 
\thanks{Based on observations carried out at Beijing
Astronomical Observatory (Xinglong, P R China) and European
Southern Observatory, La Silla, Chile}}

\subtitle{Galactic evolution and stellar depletion of lithium}

\author{Y.Q.~Chen \inst{1,2} \and P.E.~Nissen \inst{2} \and T.~Benoni \inst{2}
\and G. Zhao \inst{1}}

\offprints{P.E.~Nissen}

\institute{
Beijing Astronomical Observatory, Chinese Academy of Sciences, Beijing 100012, P R China \\
\email {cyq@yac.bao.ac.cn}
\and Institute of Physics and Astronomy, University of Aarhus, DK--8000
Aarhus C, Denmark \\
\email{pen@ifa.au.dk}
}

\date{Received 9 January 2001 / Accepted 7 March 2001}

\abstract {We present a survey of lithium abundances in
185 main-sequence field stars with $5600 \la \teff \la 6600$~K
and $-1.4 \la \feh \la +0.2$ based on new measurements of the
equivalent width of the $\lambda 6708$\,\Lione\ line in high-resolution
spectra of 130 stars and a reanalysis of data for 55 stars from
Lambert et al. (\cite{Lambert91}). 
The survey takes advantage of improved photometric and spectroscopic
determinations of effective temperature and metallicity as well as
mass and age derived from Hipparcos absolute magnitudes,
offering an opportunity to investigate the behaviour of Li as a function of
these parameters.
An interesting result from this study is the presence of a large gap
in the $\nli - \teff$ plane, which distinguishes 
`Li-dip' stars like those first identified in the Hyades cluster by
Boesgaard \& Tripicco (\cite{Boesgaard86})
from other stars with a much higher Li abundance.
The Li-dip stars concentrate on a certain mass,
which decreases with metallicity from about 1.4 $M_{\sun}$ at
solar metallicity to 1.1 $M_{\sun}$ at $\feh \simeq -1.0$. 
Excluding the Li-dip stars and a small group of lower mass stars 
with $\teff < 5900$\,K and $\nli < 1.5$,
the remaining stars, when divided into four metallicity groups,
may show a correlation between Li abundance and stellar mass.
The dispersion around the $\nli$ --mass relation is about 0.2 dex below
$\feh \simeq -0.4$ and 0.3 dex above this metallicity, which cannot
be explained by observational errors or differences in metallicity.
Furthermore, there is no correlation between the residuals of the
$\nli$-mass relations and stellar age, which ranges from 1.5~Gyr to about
15~Gyr. This suggests that Li depletion occurs early
in stellar life and that other parameters than stellar mass and metallicity
affect the degree of depletion, e.g. initial rotation velocity and/or
the rate of angular momentum loss. It cannot be excluded, however, that
a cosmic scatter of the Li abundance in the Galaxy at a given metallicity
contributes to the dispersion in Li abundance. These
problems make it difficult to determine the Galactic evolution of Li
from the data, but a comparison of
the upper envelope of the distribution of stars in the 
$\nli - \feh$ plane with recent Galactic evolutionary models
by Romano et al. (\cite{Romano99}) suggests that
novae are a major source for the Li production in the Galactic disk;
their occurrence seems to be the explanation of the steep increase of
Li abundance at $\feh \simeq -0.4$.
}
 
\maketitle

\section{Introduction}

Following the discovery by Spite \& Spite (\cite{Spite82}) of a 
uniform abundance of lithium in unevolved halo stars with $\teff > 5700$ K,
this trace element has attracted much attention, and many papers
have addressed problems of Galactic evolution and stellar depletion
of Li. The diagram of Li abundance versus metallicity for F and G stars
illustrates these problems. Below $\feh \simeq -1.4$ one sees a lithium
plateau with a very small dispersion of the Li abundance
(e.g. Spite et al. \cite{Spite96}) and perhaps a slight slope
of Li vs. \feh\ (Ryan et al. \cite{Ryan99}). Above $\feh \simeq -1.4$
the upper envelope of the distribution of stars in the $\nli$ -- $\feh$ diagram
increases from the plateau value $\log \epsilon$\,(Li)\, $\simeq 2.2$
to about 3.0 at $\feh \simeq 0.0$ (Lambert et al.
\cite{Lambert91}). In the same metallicity range 
the Li abundances show, however, an enormous variation
at a given metallicity, i.e. more than 3 dex.     

The prevalent interpretation of the  $\nli$ -- $\feh$ diagram is that the plateau
represents the primordial Big Bang \Liseven\
abundance and that the upper envelope 
of the distribution reflects the Galactic evolution of lithium.
Before adopting this explanation we should,
however, be able to explain why stars with $\feh \ga -1.4$ has such a
large dispersion of their Li abundances. Furthermore, it is not clear
if stars along the upper envelope have retained their original 
Li abundances or suffered a mild degree of Li depletion. One may also
ask if there has been a homogeneous evolution of Li in the
Galaxy as a function of \feh\ or if there is a dispersion of the Li 
abundance in the interstellar medium at a given metallicity related to
time or position in the Galaxy.

In an interesting paper on Li abundances in 81 dwarf stars, Lambert et al.
(\cite{Lambert91}) discovered that dwarf stars with $5900 < \teff < 6600$ K
tend to have a bimodal distribution in the $\nli$--\teff\ diagram (see their
Fig. 4). Stars belonging to a `high-Li' group show an increasing
abundance of Li with \teff\ and \feh . Most of the stars in the
`low-Li' group seem to have evolved from $\teff$(ZAMS) $\simeq 6600$~K,
i.e. the temperature of the `Li-dip' stars discovered in the Hyades
by Boesgaard \& Tripicco (\cite{Boesgaard86}), but a few low-Li stars have no
connection to the Li-dip. Still, they have an order of
magnitude less lithium than stars in the high-Li group
with corresponding  parameters \teff , \Mv , and \feh . 

In a comprehensive work on Li abundances of slightly evolved F-type
disk stars, Balachandran (\cite{Balachandran90}) also identified a number
of Hyades dip-like field stars, which all appear to have evolved
from a \teff(ZAMS) range of 6500 to 6800~K despite of their metallicity
differences.  As pointed out by Balachandran
this means that the characteristic mass of the Li-dip stars
decreases with metallicity.

Recently, Romano et al. (\cite{Romano99}) have studied the Galactic
evolution of Li by comparing the distribution of stars in the
$\nli -\feh$ diagram with predictions from chemical evolution models
that include several sources of 
\Liseven\ production. From a compilation of literature data they suggest
that the Li plateau extends up to $\feh \simeq -0.5$
with a steep rise of the Li abundance for higher metallicities.
In order to reproduce this trend, novae have to be included
as a dominant contributor to the Li production in the Galaxy in addition
to the contribution from AGB stars, Type II SNe and cosmic ray processes.

In order to verify and extend the interesting findings of
Balachandran (\cite{Balachandran90}), Lambert et al.
(\cite{Lambert91}) and Romano et al. (\cite{Romano99}), we have measured
the equivalent width of the $\Lione \, \lambda6708$ line in 133 main seqeuence
field stars with $5600 \la \teff \la 6600$~K and $-1.4 \la \feh \la +0.2$.
Li abundances
have been derived from a model atmosphere analysis of the data,
and stellar masses and ages have been derived from a comparison of \teff , \Mv\ 
values with stellar evolutionary tracks in the HR diagram using Hipparcos
parallaxes to determine the absolute magnitudes of the stars.
Furthermore, the Lambert et al. (\cite{Lambert91}) data have been analyzed
in the same way. The resulting
large homogeneous set of Li abundances is used to rediscuss
the distribution of stars in the $\nli - \teff$ and $\nli - \feh$
diagrams aiming at
a better understanding of the Galactic evolution of lithium and
the depletion in stars.

\section{Observations and data reductions}
\label{sec:obs}
\begin{figure}
\resizebox{\hsize}{!}{\includegraphics{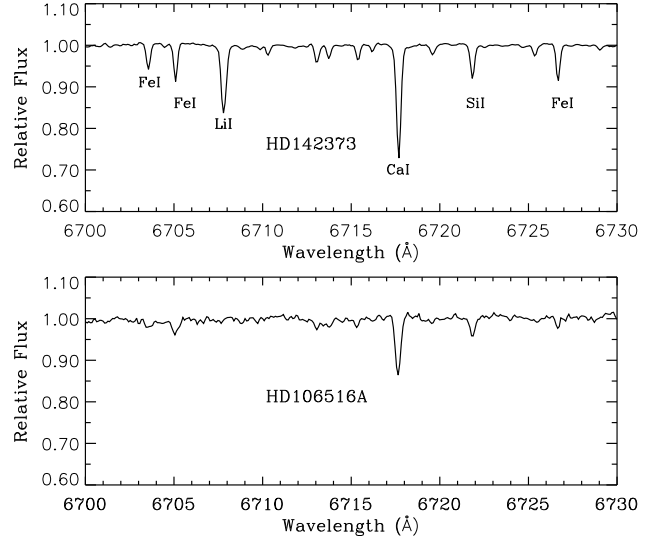}}
\caption{Representative Xinglong spectra in the region of the
$\Lione \, \lambda6708$ line for \object{HD\,142373}
($\teff = 5920 $~K, $\logg = 4.27$, $ \feh = -0.39 $ and a high $S/N \sim 400$)
with a quite strong lithium line and \object{HD\,106516A}
($\teff = 6135 $~K, $\logg = 4.34$, $ \feh = -0.71 $
and a relatively low $S/N \sim 160$) without Li detection.}
\label{fig:sptwop3}
\end{figure}

The primary set of $\Lione \, \lambda6708$ equivalent widths was obtained
from CCD spectra of 90 main-sequence stars with 
$5700 < \teff < 6600$~K and $-1.0 < \feh < +0.1$
observed with the Coud\'{e} Echelle Spectrograph at
the 2.16m telescope of Beijing Astronomical Observatory;
see Chen et al. (\cite{C2000}), who have used these spectra in connection
with a large survey of abundances of heavier elements in F and G disk 
dwarfs. The spectra cover the region 5500 - 9000~\AA\ at a resolution
of 40\,000 and have $S/N$ above 150; see Fig.~\ref{fig:sptwop3}.

The second set of equivalent widths was measured in spectra
of 28 dwarfs with $5500 < \teff < 6500$~K and $-1.4 < \feh < -0.5$ 
observed with the ESO NTT EMMI echelle spectrograph
at a resolution of 60\,000 and $S/N \ga  150$.
These spectra have previously been used by Nissen \& Schuster (\cite{Nissen97})
in a  study of the chemical composition of halo and disk stars with 
overlapping metallicities. 

The third sample consists of 15 turnoff stars
with $-0.8 < \feh < +0.2$. They were observed
in the \Lione\ line region with the 
ESO 1.4m CAT telescope and the CES spectrograph at
a resolution of 105\,000 and a very high $S/N$ of 300 to 1000.
These spectra were primarily obtained for a study of the lithium isotope
ratio and some of them have been analyzed by Nissen et al.
(\cite{Nissen99}), who detected \Lisix\ in two stars 
(\object{HD\,68284} and \object{HD\,130551}) at a level corresponding to
$\sixseven \simeq 0.05$. Here we use the equivalent width of the \Lione\
line to derive the total Li abundance of the stars.

Due to the complications by measuring reliable equivalent widths and
making a proper model atmosphere analysis of
double-lined spectroscopic binaries such stars were excluded from the
three samples. Proven or suspected single-lined spectroscopic binaries 
were, however, retained. According to Chen et al. (\cite{C2000}) such
stars do not appear to have any abundance anomalies of the heavier elements.
As discussed later there are, however, indications that the Li abundance
may be peculiar for such SB1 stars.

The spectra were reduced using standard MIDAS 
(Xinglong data) and IRAF (ESO data) routines for order definition,
background correction, flatfielding,
extraction of echelle orders, wavelength calibration and
continuum fitting. At the resolution of the Xinglong and ESO NTT spectra,
the profile of the \Lione\ doublet is well approximated by a Gaussian function 
despite of its inherent asymmetry. Hence, the equivalent width of the
\Lione\ line was measured by Gaussian fitting, which has the advantage that the 
weak $\Feone \, \lambda$6707.44 line in the blue wing
of the Li line gives no significant contribution. For the high resolution,
high $S/N$ spectra from
the ESO CES the equivalent width was measured both by Gaussian fitting and
by direct integration excluding the blending \Feone\ line. The two sets
of data agree within $\pm 1$\,m\AA .

\begin{figure}
\resizebox{\hsize}{!}{\includegraphics{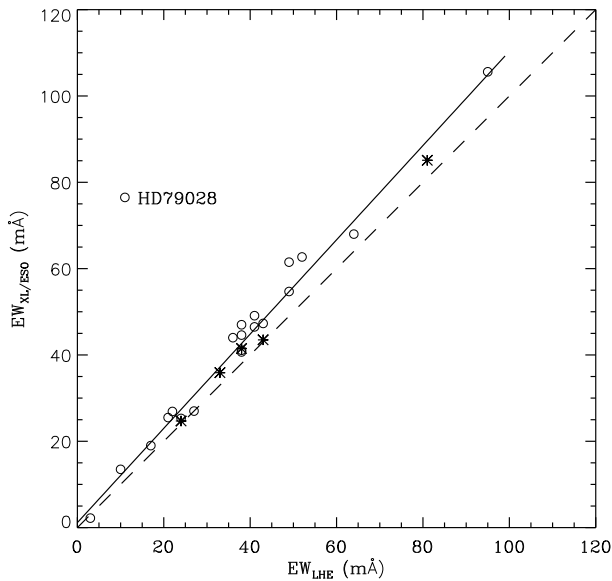}}
\caption{Equivalent widths of the $\Lione \, \lambda6708$ line
measured from Xinglong (open circles) or ESO (asterisks)
spectra compared to equivalent widths given by
Lambert et al. (\cite{Lambert91}).}
\label{fig:comewp3}
\end{figure}

Lambert et al. (\cite{Lambert91}) measured their equivalent widths
from spectra obtained  at McDonald Observatory with  
the Coud\'{e} Spectrographs of the 2.1 and 2.7m telescopes.
The resolution of these spectra is around 35\,000 and $S/N \sim 150$.
Twenty stars are in common with the Xinglong sample and 5 stars
with the ESO sample. A comparison of the equivalent widths 
is shown in Fig.~\ref{fig:comewp3}. Obviously, Lambert et al.'s
EWs are systematically lower than the Xinglong and ESO data.
The reason for this discrepancy is unclear. A linear least square fit gives
\begin{eqnarray}
EW_{\mathrm{XL}}=1.09 \left(\pm0.03\right) EW_{\mathrm{LHE}}+1.12\left(\pm1.26\right) \,\,
(m{\mbox{\AA}})
\label{eq:ewcom}
\end{eqnarray}
where the deviating star \object{HD\,79028} has been been excluded
from the regression. We suspect that this
star was misidentified by Lambert et al. because
there is a very good agreement between Xinglong EWs measured by
Chen et al. (\cite{C2000}) and those of Edvardsson et al. (\cite{Edvardsson93})
for 26 metal lines in common.

The relation between the Lambert et al. equivalent widths and those 
based on the Xinglong and ESO spectra appears to be well defined; the
rms scatter around the line shown in Fig.~\ref{fig:comewp3} is less 
than 3 m\AA\ (excluding \object{HD\,79028}). Hence, we have rescaled
the Lambert et al. data according to Eq. 1 to get a homogeneous set
of equivalent widths of $\Lione \, \lambda6708$.

\section{Analysis}
\label{sec:analysis}

Effective temperatures were determined 
from the Str\"{o}mgren $(b-y)$ color index using the
calibration of Alonso et al. (\cite{Alonso96}).
For the large majority of stars the gravities were determined via Hipparcos
parallaxes (ESA \cite{ESA97}) as described by Chen et al. (\cite{C2000}).
For the more distant stars 
we adopt a spectroscopic gravity obtained by requiring
that \Feone\ and  \Fetwo\ lines provide the same iron abundance.
Metallicities were taken from Chen et al. (\cite{C2000}), 
Nissen \& Schuster (\cite{Nissen97}), and Edvardsson et al. 
(\cite{Edvardsson93}). For a few stars not included
in these sources, \feh\ was derived from the equivalent widths of the 
$\lambda6703$ and $\lambda6705$ \Feone\ lines given by Lambert et al.
using ``solar'' oscillator strengths. Microturbulences were taken from 
Chen et al. (\cite{C2000}) or calculated from the empirical relation found by
Edvardsson et al. (\cite{Edvardsson93}). 
The uncertainties of the parameters are: $\sigma (\teff ) = 70$~K, 
$\sigma (\logg ) = 0.1$~dex, $\sigma (\feh ) = 0.1$ and $\sigma
(\xi )=0.3 \kmprs $.


The model atmospheres were interpolated from a grid of 
plane-parallel, LTE models computed with the MARCS code
by Bengt Edvardsson (Uppsala). The corresponding analysis program, 
SPECTRUM, was used to calculate equivalent widths of the \Lione\ line
as a function of the Li abundance with wavelengths and oscillator
strengths of the Li doublet components taken from
Sansonetti et al. (\cite{Sansonetti95})
and Yan \& Drake (\cite{Yan95}), respectively (see Smith et al.
\cite{Smith98}, Table 3). 
The contribution of the $^6 \mathrm{Li}$ isotope was assumed to be
negligible.
Li abundances were then obtained by requiring that the theoretical 
equivalent widths should match the observed ones. 

The uncertainty in the Li abundance, resulting from errors of the
equivalent widths, is around 0.05 dex for stars with
EW $\sim$ 10-120 m\AA\ increasing rapidly for stars with weaker lines.
Changes of the Li abundance due to errors in the model atmosphere 
parameters are 0.058 dex for $\Delta T=+70$~K, $-$0.001
dex for both $\Delta \log g=+0.1$ and $\Delta \feh=+0.1$,
$-0.002$ for $\Delta \xi_{t}=+0.3 \kmprs$. These numbers were derived 
for \object{HD\,142373} with $\teff=5920$ K,
$\logg=4.27$, $\feh=-0.39$ and a \Lione\ equivalent width of 49 m\AA , but
are representative for the whole sample.
For the Lambert et al. sample our derived Li abundances agree with their
original values and those derived by  Romano et al. (\cite{Romano99})
within 0.1 dex. The differences are mainly due to different ways of determining
\teff\ and our rescaling of the EWs published by Lambert et al.

Finally, non-LTE
corrections were applied to the derived Li abundances based on the 
work by Carlsson et al. (\cite{Carlsson94}), who studied non-LTE
formation of the $\Lione \, \lambda6708$ line as a function of effective
temperature,
gravity, metallicity and Li abundance. For the present sample of stars, the 
largest correction (LTE -- non-LTE) is $0.12$ dex for the hottest and
most Li-rich stars; the correction decreases with
temperature and becomes slightly negative for the cool Li-poor stars.

Atmospheric parameters and Li abundances are presented in 
Table~\ref{tb:all}, which also includes the absolute magnitude derived from 
the Hipparcos parallax (ESA \cite{ESA97}). In a few cases where the
Hipparcos parallax is not available or has a low 
relative accuracy the Str\"{o}mgren $c_{1}$ index has been used to
determine $\Mv$ (see
Edvardsson et al. \cite{Edvardsson93}). Stellar masses and ages derived
from evolutionary tracks of VandenBerg et al. (\cite{VandenBerg99})
(see Sect. 4.2) are also given in Table~\ref{tb:all}.

\section{Discussion}
\subsection{Li vs. effective temperature}

\begin{figure}
\resizebox{\hsize}{!}{\includegraphics{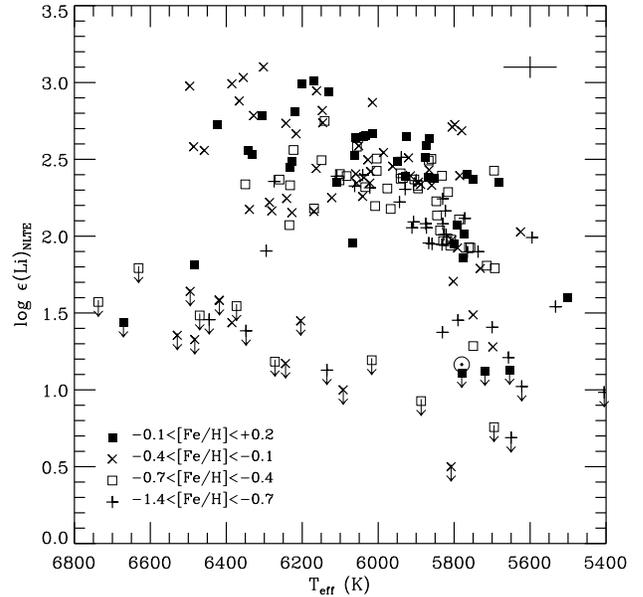}}
\caption{Lithium abundance versus effective temperature for stars in the
present survey. Upper limits of the Li abundance are indicated with 
downward directed arrows.} 
\label{fig:LiTeff}
\end{figure}

Fig.~\ref{fig:LiTeff} shows the Li abundance vs. $\teff$ for all stars
in the present survey.
As seen, the stars tend to separate into two groups, one with  high and
one with low Li abundances. Although there is a significant spread of
the Li abundance within each group, a gap of at least 1.0 dex is seen for
$\teff > 5900$\,K suggesting that a special and very rapid
Li-depletion mechanism is operating in the interior of the low-Li stars.

In the low-Li group almost all abundances derived are upper limits,
and the apparent linear correlation between the Li abundance and \teff\      
simply reflects the observational detection limit, EW$\simeq 3$\,m\AA ,
which corresponds to a Li abundance of 1.0 dex
at $\teff=5800$~K, 1.3 dex at $\teff=6200$~K and 1.6 dex
at $\teff=6600$~K. As discussed in the next section,
most of the low-Li stars with $\teff > 5900$\,K
have evolved from an effective temperature range corresponding
to that of the Hyades Li-dip stars.

In the high-Li region, there is a spread of 1.0 dex in Li abundance 
at a given $\teff$.  Metallicity seems to be responsible
for the main part of the scatter based on the fact that the upper
envelope consists of stars with $\feh >-0.4$, while nearly all 
stars with $\feh <-0.4$ have a Li abundance lower than
2.5 dex.
This result, coupled with the fact that there is no large Li abundance
differences among stars with $-1.4 < \feh < -0.6$  and $\teff > 5900$~K,
suggests a significant increase of the Galactic production of lithium
around $\feh \sim -0.4$.

In the $\teff>  5900$~K range, the mean Li abundance of stars in
the high-Li region shows a decrease with temperature for
$\feh > -0.4$ while it is  nearly $\teff$ independent 
for $\feh <-0.4$. In both metallicity groups a steep decline
of the Li abundance begins at $\teff \simeq 5900$~K. Furthermore,
a large scatter in Li abundance is apparent in the range 
$5600 \la \teff \la 5900$
with a tendency of a bimodal distribution. The Sun with a 
Li abundance of $\nli = 1.16$ (Steenbock \& Holweger \cite{Steenbock84})
belongs to the low-Li group.

\subsection{Stellar masses and ages}

In order to study the behaviour of Li as a function of stellar
mass and age, the positions of the stars
in the $\Mv$ -- $\log \teff$ diagram were compared to
the evolutionary tracks of VandenBerg et al. (\cite{VandenBerg99}).
Masses and ages were derived as described by Chen et al. (\cite{C2000}).
The typical error of the mass is 0.03-0.06~$M_{\odot}$. For most
of the stars the age could be determined with an error of
15-20~\%, but a number of stars situated close to the ZAMS or in the hook
region of the evolutionary tracks
have much larger age errors. For these stars no age
is given in Table 1.

\begin{figure*}
\centering
\resizebox{\hsize}{!}{\includegraphics{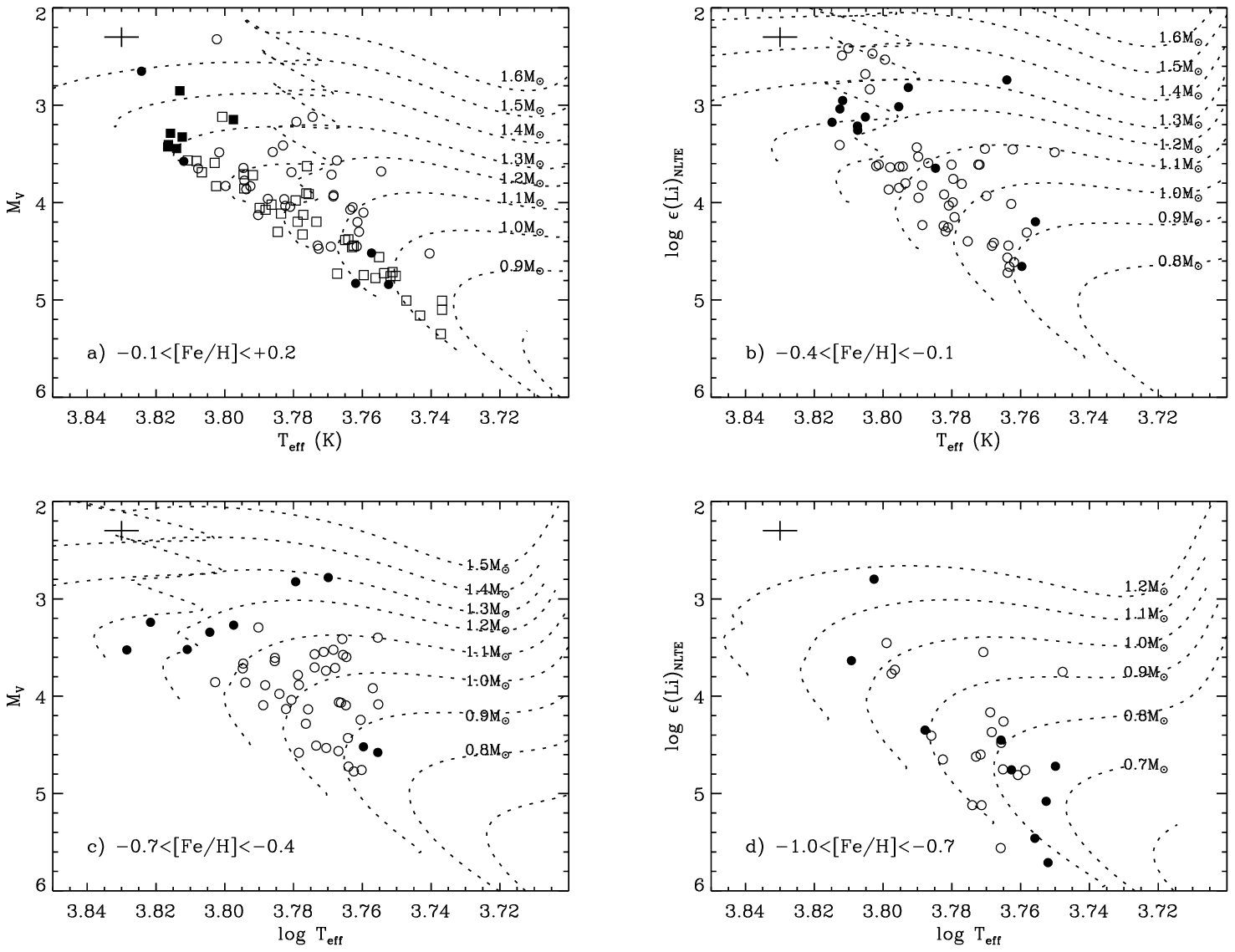}}
\caption{Stellar positions in the \Mv - \logteff\ diagram
divided according to metallicity and compared to
evolutionary tracks of VandenBerg et al. (\cite{VandenBerg99})
corresponding to:
{\bf a}) $\feh = +0.12$,
{\bf b}) $\feh = -0.20$, 
{\bf c}) $\feh = -0.53$, and
{\bf d}) $\feh = -0.83$.
The filled circles indicate stars in the low-Li group 
and open circles stars in the high-Li group. 
In a) Hyades stars have been added and 
are plotted with squares instead of circles.}
\label{fig:LiHR}
\end{figure*}

Fig.~\ref{fig:LiHR} shows the position of the stars in the
$\Mv$ -- $\log \teff$ diagram for four metallicity ranges.
In panel a), that contains the most
metal rich stars, 46 Hyades stars are added.
They were selected from the papers of
Boesgaard \& Tripicco (\cite{Boesgaard86}),
Boesgaard \& Budge (\cite{Boesgaard88}) and
Thorburn et al. (\cite{Thorburn93}) with the condition that they should
lie in the same \teff\ range as our program stars, and excluding known
binaries. Str\"{o}mgren photometry of the Hyades stars was taken from
Crawford \& Perry (\cite{Crawford66}) and Hipparcos parallaxes
from Perryman et al. (\cite{Perryman98}). The effective temperatures
were determined in the same way as for the program stars, and the
derived (non-LTE) Li abundance 
is based on the Li-Fe blended EWs given in the three papers
taking into account the contribution
of the \Feone\ $\lambda 6707.44$ blending line using
log $gf = -2.29$ as derived from an analysis of the solar flux spectrum.

As seen from Fig.~\ref{fig:LiHR}, stars in the low-Li region (filled circles) 
are concentrated either in the high mass or the low mass region
at a given metallicity. The high-mass, low-Li stars probably
suffer from the same depletion mechanism that acts on the Hyades.
Like the Hyades Li-dip stars they are grouped in a quite narrow mass range.
This is particular clear from panels b) and c). Furthermore,
the mass of the Li-dip stars seems to be 
metallicity dependent with lower mass for the more metal-poor stars.
According to Fig.~\ref{fig:LiHR}, the critical mass 
is about $1.4 M_{\sun}$ at $\feh \simeq 0.1$, $1.3 M_{\sun}$ 
at $\feh \simeq -0.20$, $1.2 M_{\sun}$ at $\feh \simeq -0.5$, and
$1.1 M_{\sun}$ at $\feh \simeq -0.9$.
We also note that Li-dip stars can be found over a range
of effective temperatures as
clearly shown by panels b) and c), where stars with $\teff \sim  6600$
K and $\teff \sim 5700$ K both show the Li-dip depletion. Therefore,
Li-dip stars in the field can not be distinguished by the
temperature range alone. Instead, stellar mass is the key
parameter that decides if a star suffers from the Li-dip
depletion. 

In the low-mass region there is a  tendency that the lowest
Li abundances occur among stars with the smallest masses,
but there is no clear separation between high and low-Li stars.
Apparently, the Li abundance is very sensitive to stellar mass 
and/or metallicity in this part of the HR diagram.

Two low-Li stars have intermediate masses, namely 
\object{HD\,80218} in panel b), and \object{HD\,106516A} in panel d).
\object{HD\,80218} was recently shown to be an astrometric
binary by Gontcharov et al. (\cite{Gontcharov00}).
\object{HD\,106516A} (HR~4657)
is a single-lined spectroscopic binary with an unusual large
rotational velocity ($V$sini = 6.8~\kmprs , Fuhrmann \& Bernkopf
\cite{Fuhrmann99}) for its effective temperature (\teff = 6135~K)
and metallicity ($\feh = -0.71$). Fuhrmann \& Bernkopf
suggest that \object{HD\,106516A}  is a blue straggler and 
link the absence of Li to this special class of objects.

Lambert et al. (\cite{Lambert91}) also classified \object{HD\,106516A}
as a Li-poor star unrelated to both the Li-dip stars
and the low-mass stars. They suggested six other stars as belonging
to this `intermediate-mass' group of Li-poor stars. None of these
cases are, however, confirmed by our work. Based on the more
accurate Hipparcos distances \object{HD\,95241} (\object{HR\,4285)},
\object{HD\,159332} (\object{HR\,6541)} and \object{HD\,219476} 
clearly belong to the Li-dip stars. 
\object{HD\,30649} and \object{HD\,143761} (\object{HR\,5968)}
belong to the low-mass, Li-poor group. Finally,
\object{HD\,79028} (\object{HR\,3648)} is the star probably
misidentified by Lambert et al. As discussed in Sect. 2, we 
measure a much larger equivalent width of the Li line, and hence
the star belongs to the high-Li group.
Altogether, we conclude that `intermediate-mass' Li-poor stars are rare
and may be binaries.

\subsection{Li vs. metal abundance}

\begin{figure}
\resizebox{\hsize}{!}{\includegraphics{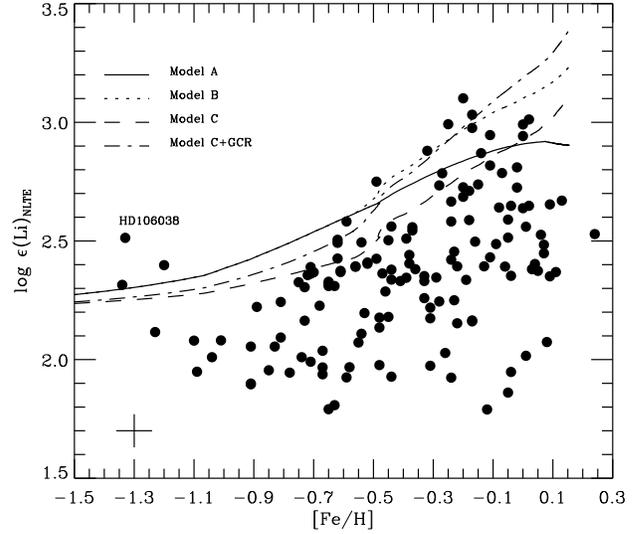}}
\caption{Lithium abundances as a function of metallicity  for stars
in the high-Li group (see text for the definition).
Theoretical predictions
from Romano et al. (\cite{Romano99}) are plotted.}
\label{fig:LiFe}
\end{figure}

In agreement with previous studies, a plot of $\nli$ as a
function of $\feh$ (see Fig.~\ref{fig:LiFe}) indicates 
a large spread in Li abundance
at a given metallicity, and the abundance range tends to increase with
higher metallicity. The upper envelope of the distribution of $\nli$
for  $\feh<-0.2$  is compatible with the the relation
given by  Lambert et al.  (\cite{Lambert91})
as derived from the Rebolo et al. (\cite{Rebolo88}) data.
Above $\feh \simeq -0.2$ our sample is, however, not reaching
such high Li abundances as the Rebolo et al. sample. This is partly
due to the neglect of non-LTE correction by Rebolo et
al., which leads to an overestimate of the Li
abundance by about 0.1 dex for stars with the highest Li abundances.
Furthermore, we note that stars with the highest Li abundances in this
metallicity range in Rebolo et al. (\cite{Rebolo88})
all have $\teff>6500$~K. These rather young, metal-rich stars are not
represented in our sample of stars.

In order to show the upper envelope more clearly, Fig.~\ref{fig:LiFe}
includes only stars belonging to the high-Li group
defined as stars having $\nli > 2.0$ for $\teff > 5900$~K
and $\nli > 1.75$ for $\teff \le 5900$~K. As seen from Fig. 3 this
definition excludes all Li-dip stars and the group of low-Li stars
for $\teff \le 5900$~K.

The distribution of stars in Fig.~\ref{fig:LiFe} 
is compared to recent models for the Galactic evolution
of Li by Romano et al. (\cite{Romano99}).
Their model A includes contributions to the Li production from low-mass
AGB stars (C-stars), high-mass AGB stars and SNeII. As seen this model
gives a very poor fit to the upper envelope. In model B, Li production
by novae is included, which improves the fit at higher metallicities,
but the models fails in predicting a too high Li abundance at 
metallicities around $\feh \sim -0.7$. Model C has no contribution from
low-mass AGB stars and the yields of high-mass AGB stars and SNeII are
reduced by a factor of two, whereas the contribution from novae is retained.
This model provides a better fit of the upper envelope at low metallicities
by predicting only a weak increase of the Li abundance from $\feh = -1.5$
to $-0.5$ and then a steep increase, but the model fails to account for 
the high Li abundance ($\sim 3.0$ dex) for stars with $-0.3 < \feh < -0.1$.
Finally, model C+GCR also includes the contribution to Li production from
cosmic ray processes. This model gives the best overall fit especially when
one takes into account that all stars with $\feh > -0.1$ probably
have suffered some
Li depletion, as suggested by the high (3.3 dex) Li abundance in meteorites
and in young stellar clusters like the Pleiades.  

One low metallicity star (\object{HD\,106038} at $\feh = -1.33$ 
and $\nli = 2.52$) lies well above
the Li evolutionary curves, and also above the `Spite plateau'
for metal-poor halo stars. As shown by Nissen \& Schuster (\cite{Nissen97})
this star has very peculiar abundances; [Si/Fe] $\simeq 0.6$,
[Ni/Fe] $\simeq 0.2$, [Y/Fe] $\simeq 0.4$, and [Ba/Fe] $\simeq 0.5$.
Most likely, the atmospheric composition of the star has been changed
by mass transfer from an evolved component in its AGB phase causing
the enrichment in Li and the $s$-process elements. The enrichment
in Si and Ni is, however, difficult to explain.

\begin{figure*}
\centering
\resizebox{\hsize}{!}{\includegraphics{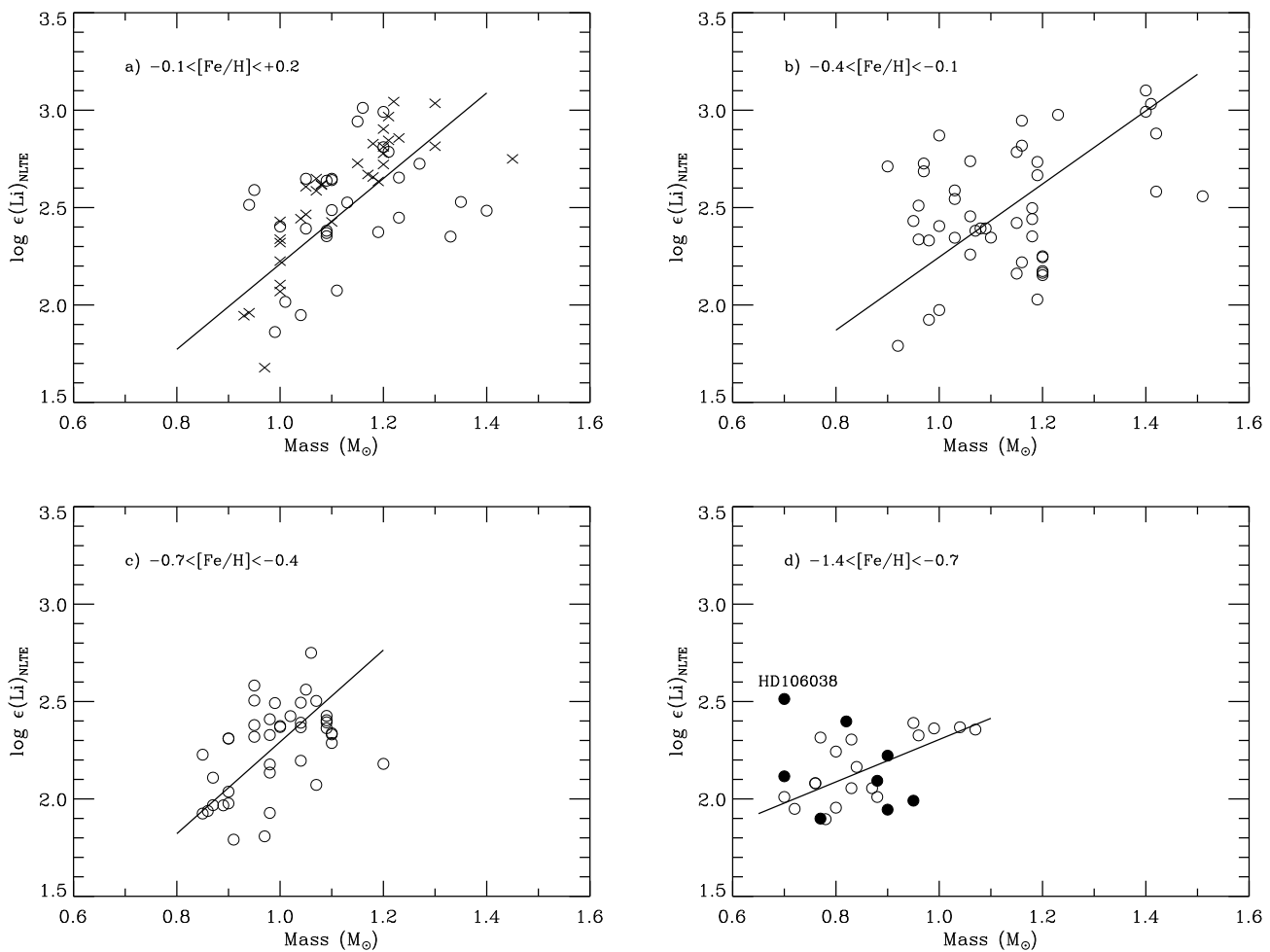}}
\caption{Li abundance versus stellar mass for `high-Li' stars
divided into four metallicity groups. In {\bf a)} Hyades stars outside the 
Li-dip have been added and are plotted with crosses. In {\bf d)} stars
with halo kinematics have been plotted with filled circles. The rest of the
stars have disk kinematics.}
\label{fig:LiFemass}
\end{figure*}

\subsection{Li vs. stellar mass and age}
It is evident from the $\nli$ -- $\feh$ diagram (Fig.~\ref{fig:LiFe})
that only a few stars lie along the Li evolutionary curves predicted from
the models of Romano et al. (\cite{Romano99}). The canonical interpretation
is that Li has been destroyed in the majority of the stars due to
reactions with protons at the bottom of the outer convection zone of the
star, i.e. at $T > 2.5 \, 10^6$~K. As the depth of the convection zone
depends primarily on stellar mass a correlation between
Li abundance and mass is then expected for a given metallicity. To investigate
this, we have divided the high-Li stars (i.e. those included in
Fig.~\ref{fig:LiFe}) into four
metallicity intervals and plotted the Li abundance as a function of
mass (see Fig.~\ref{fig:LiFemass}).  In panel a), Hyades stars outside the
Li-dip region have been added.

As seen from Fig.~\ref{fig:LiFemass} there may indeed be a correlation
between Li abundance and mass. 
Excluding the Hyades and the peculiar star (\object{HD\,106038}
straight lines have been fitted to the data 
taking into account errors in both coordinates.
The slope of the fitted lines is about the same in panels
a), b) and c), $\Delta \nli / \Delta$Mass $\simeq 2$.
For the most metal-poor group, panel d), the slope is
$\simeq 1$ but the correlation is marginal.
It is also seen that the slope defined by the Hyades stars
agrees well with that of the field stars and that the Hyades lie only
slightly above the mean relation for the field stars
despite of their younger age.

The dispersion around the fitted lines in Fig.~\ref{fig:LiFemass} is, 
however, much
larger than expected from the estimated errors of $\nli$ and the stellar
mass. From $\sigma \nli \simeq 0.10$ and $\sigma$(Mass) $\simeq 0.05 M_{\sun}$
we would expect a dispersion in the $\nli$ direction of 0.16 dex in 
panels a), b) and c) and 0.11 dex in panel d). The actual dispersion
around the fitted lines is 0.32 dex in panels a) and b), 0.22 dex in c)
and 0.14 dex in d). Furthermore, there is no significant
correlation between the residuals of the fits and the stellar age
(see Fig. 7) or 
the metallicity variation within each group. The most likely explanation
is that the depletion of Li depends on other parameters
than stellar mass, age and metallicity  such as the initial rotational
velocity of the star and/or the rate of angular momentum loss
during its evolution.
The same conclusion has been reached from studies of Li abundances
of open cluster stars, e.g. the Hyades (Thorburn et al. \cite{Thorburn93}) and
M67 (Jones et al. \cite{Jones99}) as well as from studies of
Li abundances in upper main sequence stars (Balachandran
\cite{Balachandran90}), solar-like stars (Pasquini et al. \cite{Pasquini94})
and subgiants (Randich et al. \cite{Randich99}).

\begin{figure}
\resizebox{\hsize}{!}{\includegraphics{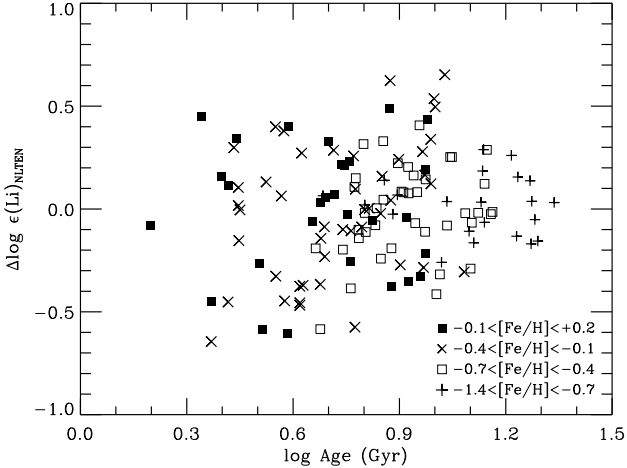}}
\caption{The residuals of $\nli$ with respect to the $\nli$-mass fits
in Fig.~\ref{fig:LiFemass} as a function of the
logarithmic stellar age. The four metallicity groups are shown by
different symbols.}
\label{fig:res}
\end{figure}

The fact that there is no significant correlation between the residuals
of the $\nli$-mass relations and stellar age is interesting.  As the age for
the present sample of stars ranges from 1.5~Gyr to about 15 Gyr, the lack
of correlation suggests that the main part of Li depletion occurs rather
early in the stellar life, i.e. at ages $\la 1.5$~Gyr. In a review of Li
abundances in open cluster stars, Pasquini (\cite{Pasquini00}) reached a
similar conclusion: most of the Li depletion occurs for stellar ages between
$\sim 100$ and $\sim 600$~Myr.

There may, however, also be a contribution to the Li 
dispersion arising from a cosmic scatter of Li in the Galaxy
at a given metallicity. To investigate this we have searched for
correlations between Li abundances and stellar kinematics, but 
the outcome was negative. In the lowest metallicity group 
($-1.4 < \feh < -0.7$) eight stars have halo kinematics, i.e. elongated
Galactic orbits and low values of the velocity component in the direction
of Galactic rotation, $V_{\rm LSR} \la -200$~\kmprs .
Furthermore, Nissen \& Schuster (\cite{Nissen97}) have shown that these
`metal-rich' halo stars have lower ratios between the
$\alpha$-elements and Fe than disk stars in the same metallicity range.
As seen from Fig.~\ref{fig:LiFemass} panel d) there is, however, no systematic
differences between the Li abundance of stars with 
halo and disk kinematics, respectively. Still, we cannot exclude
variations in Li/Fe on Galactic scales that are not reflected in
the kinematics of the stars. The fact that Be abundances of
disk main sequence stars with \teff\ between 5700 and 6400~K
show a large spread (Boesgaard \& King \cite{Boesgaard93}) suggests
inefficient mixing of light elements produced by cosmic rays
considering that beryllium is more robust against stellar destruction
than lithium and is unlikely to be depleted in the temperature
range mentioned.

\begin{figure}
\resizebox{\hsize}{!}{\includegraphics{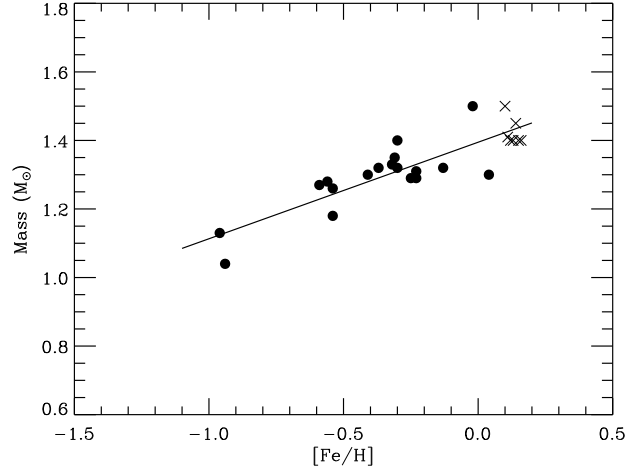}}
\caption{The dependence of stellar mass on metallicity for Li-dip stars. 
Crosses refer to Hyades stars and filled circles to field stars.}
\label{fig:LidipFeM}
\end{figure}

The low-Li stars (not included in Fig.~\ref{fig:LiFemass}) can be
divided into two classes: lower mass stars with $\nli \la 1.5$ to which
the Sun belongs, and higher mass stars originating from a Hyades-like
Li dip. For the lower mass, Li-poor stars we note that adding them
to Fig.~\ref{fig:LiFemass}
greatly increases the dispersion in Li at the lower mass end of the
plots.
For the Li-dip stars a tight correlation
between metallicity and mass is found  (see Fig.~\ref{fig:LidipFeM}).
A least squares fit to the data gives the following relation
\begin{eqnarray}
M = (1.40 + 0.28 \feh ) M_{\sun}
\end{eqnarray}
with a dispersion that can be fully explained by the estimated errors
in \feh\ and mass. Clearly, the special depletion of
Li associated with the Hyades-like Li-dip occurs within a narrow
mass range at a given metallicity. 

It is noted that an extrapolation of Eq. 2 leads to a Li-dip 
mass of about $1 M_{\sun}$ at $\feh \simeq -1.5$. This is well above
the maximum turnoff mass of halo stars ($\sim 0.8 M_{\sun}$)
suggesting that Li-dip stars are unlikely to occur among halo stars
at their typical metallicity of $\feh \simeq -1.5$.

\section{Conclusions}
On the basis of high precision determinations of Li abundance
and stellar parameters for a sample of
185 main-sequence field stars with $5600 \la \teff \la 6600$~K
and $-1.4 \la \feh \la +0.2$, two kinds of Li depletion emerge:
the as yet unknown process responsible for the Li-dip stars,
and the mass dependent depletion of Li, which occurs near the
bottom of the convective zone due to destruction of Li by reactions
with protons. The group of Li-dip stars is clearly separated from
the high-Li stars in
the $\nli - \teff$ diagram with a gap of at least 1 dex in the
Li abundance.

As shown in Fig.~\ref{fig:LidipFeM} there is a well defined relation between
the mass of the Li-dip stars and their metallicity confirming 
previous findings by
Balachandran (\cite{Balachandran90}) and Randich et al. (\cite{Randich99}).
The derived mass-\feh\ relation, Eq. (2), may be used as a constraint
on the various stellar mixing theories suggested as an explanation
of the Li-dip stars: microscopic diffusion,
mass loss, meridional circulation, turbulence or rotational braking;
see e.g. the discussion by Balachandran (\cite{Balachandran95}), 
the study of Talon \& Charbonnel (\cite{Talon98}), and the recent review
by Deliyannis et al. (\cite{Deliyannis00}).

Excluding the Li-dip  stars, a correlation between Li abundance
and stellar mass is found (see Fig.~\ref{fig:LiFemass}), which 
is probably due to an increasing degree of Li depletion as the
outer convection zone deepens with decreasing stellar mass.
The dispersion around the Li-mass relation is, however, much
larger than expected from observational errors and the residuals are
not correlated with stellar age, which for the present sample of stars
ranges from  $\sim 1.5$~Gyr to about 15~Gyr. This suggests that the
main part of Li depletion occurs when the star is young and that
other parameters in addition to stellar mass and metallicity
affect the degree of depletion, e.g. initial rotation velocity, 
rate of angular momentum loss or magnetic field strength.
Binarity may also play a role for the degree of Li depletion,
as exemplified by the two `intermediate-mass' Li-poor stars
\object{HD\,80218} and \object{HD\,106516A}, and although we have
excluded double-lined spectroscopic binaries from our sample,
a significant fraction of the stars may be binaries with a
low mass component.
Furthermore, it cannot be excluded that
a cosmic scatter of the Li abundance in the Galaxy at a given metallicity
contributes to the dispersion in Li abundance, although we have not
found any correlation between stellar kinematics and Li abundance.

If the upper envelope of the $\nli - \feh$ distribution is assumed to
represent the Galactic evolution of Li, then the Galactic evolutionary
models of Romano et al. (\cite{Romano99}) suggest that 
novae, massive AGB stars, SNeII and Galactic cosmic rays all
contribute to the production of Li. In particular, novae seem to
provide a major contribution to the Li enrichment and to be
responsible for the steep increase of the Li abundance at $\feh \simeq -0.4$.
It is, however, doubtful if stellar Li abundances along the 
upper envelope of the $\nli - \feh$ distribution, based on samples
of long-lived F \& G stars in this and other works, really represent
the Galactic evolution of Li given that there may be a dispersion
of Li in the interstellar medium at a given metallicity and that
no stars reach the meteoritic Li abundance at solar metallicity.
To progress on this problem one will probably need a sample of
stars that is an order of magnitude larger than the present sample.
Furthermore, it would be interesting to observe the spectra of these stars
with higher resolution, say $R \simeq 100\,000$, and very high S/N
in order to be able to estimate stellar
rotation velocities and magnetic field strengths from the profiles
of Zeeman-insensitive and Zeeman-sensitive spectral lines. This would enable
a study of possible correlations between the degree of Li depletion
and stellar rotation as well as magnetic field strength.

\section*{Acknowledgements}
This research was supported by the Danish Research Academy,
the Danish Daloon Foundation, the Chinese NSF, and NKBRSF G1999075406.

\begin{table}
\caption[ ]{Atmospheric parameters, absolute magnitude, mass, age,
equivalent width of the $\lambda 6708$~Li line,
and non-LTE Li abundance. For the majority of stars $\Mv$
has been derived from Hipparcos parallaxes but in a few cases (marked by 
:) a photometric $\Mv$ derived from the $c_{1}$ index is given. 
Most stellar ages have errors of $\simeq$20\% but those marked with `:' are 
more uncertain, and for stars close to the ZAMS no age is given}
\label{tb:all}
\setlength{\tabcolsep}{0.10cm}
\begin{tabular}{lrrrllrrr}
\noalign{\smallskip}
\hline
\noalign{\smallskip}
Star & $\teff$ & $\logg$ & \ffe & $\Mv$ & Mass & Age & E.W.& $\nlia$ \\
HD   & K     &       &      &     &$M_{\odot}$ & Gyr  &  $\mA$ & (Li)   \\
\noalign{\smallskip}
\hline
\noalign{\smallskip}
   400 &6122&4.13&$-$0.23&3.59 & 1.20&  4.3  & 26.9   & 2.25   \\
   693 &6163&4.11&$-$0.38&3.53 & 1.18&  4.8  & 38.2   & 2.44   \\
  2454 &6418&4.09&$-$0.37&3.26 & 1.32&  3.3:   & $<$4.0     &$<$1.60     \\
  3454 &6056&4.29&$-$0.59&4.13 & 0.95&  9.1  & 58.6   & 2.58   \\
  3567 &6041&4.01&$-$1.20&4.07:& 0.82& 13.7  & 45.4   & 2.40   \\
  4614 &5806&4.33&$-$0.31&4.57 & 1.00&  8.0  & 24.0   & 1.97   \\
  4813 &6146&4.34&$-$0.15&4.23 & 1.06&  3.8  & 71.0   & 2.74   \\
  5750 &6223&4.21&$-$0.44&3.86:& 1.05&  6.0  & 45.7   & 2.56   \\
  6834 &6295&4.12&$-$0.73&3.45 & 1.04&  6.3  & 10.3   & 1.90   \\
  6840 &5860&4.03&$-$0.45&3.71:& 1.07&  7.2  & 65.8   & 2.50   \\
  7439 &6419&4.13&$-$0.32&3.21 & 1.33&  2.8:    & $<$4.0    & $<$1.60     \\
  7476 &6486&3.91&$-$0.24&2.49 & 1.42&  2.6: & 32.8   & 2.58   \\
  9826 &6110&4.13&   0.09&3.48 & 1.33&  3.3  & 33.9   & 2.35   \\
 10307 &5776&4.13&$-$0.05&4.45 & 0.99&  9.1  & 19.7   & 1.86   \\
 11007 &6027&4.20&$-$0.16&3.61 & 1.18&  4.9  & 51.7   & 2.50   \\
 11592 &6232&4.18&$-$0.41&3.66 & 1.10&  5.5  & 28.0   & 2.33   \\
 13555 &6366&4.03&$-$0.32&2.84 & 1.42&  2.8: & 71.0   & 2.88   \\
 15335 &5785&3.92&$-$0.22&3.45 & 1.09&  7.0  & 57.9   & 2.39   \\
 15798 &6385&3.92&$-$0.25&2.68 & 1.40&  2.8: & 85.1   & 2.99   \\
 16673 &6170&4.35&   0.02&4.13 & 1.16&  2.2  &112.8   & 3.01   \\
 16895 &6219&4.27&$-$0.02&3.86 & 1.20&  2.5  & 74.3   & 2.81   \\
 17288 &5700&4.38&$-$0.88&5.46:& 0.80&       &  8.3   & 1.41   \\
 17820 &5750&4.11&$-$0.69&4.52:& 0.83& 16.4  &  5.6   & 1.28   \\
 18768 &5695&3.91&$-$0.62&3.40 & 1.09&  6.8  & 67.7   & 2.43   \\
 19373A&5867&4.01&   0.03&3.94 & 1.09&  5.7  & 51.2   & 2.38   \\
 22484 &5915&4.03&$-$0.13&3.61 & 1.08&  6.3  & 49.1   & 2.39   \\
 22879 &5790&4.28&$-$0.84&4.76 & 0.78& 18.8  &  7.8   & 1.45   \\
 24339 &5810&4.20&$-$0.67&4.43:& 0.86& 14.4  & 21.9   & 1.94   \\
 24421 &5986&4.10&$-$0.37&3.81 & 1.03&  7.9  & 59.7   & 2.54   \\
 25173 &5867&4.07&$-$0.62&3.52 & 0.99&  7.9  & 63.3   & 2.49   \\
 25457 &6162&4.28&$-$0.11&3.95 & 1.16&  3.5  &104.5   & 2.95   \\
 25704 &5765&4.12&$-$0.91&4.81 & 0.78& 18.7  & 21.6   & 1.90   \\
 25998 &6147&4.35&$-$0.11&3.82 & 1.16&  4.2  & 85.3   & 2.82   \\
 28620 &6101&4.08&$-$0.52&3.61 & 1.09&  6.1  & 38.2   & 2.40   \\
 30649 &5695&4.24&$-$0.51&4.58 & 0.88& 14.5     & $<$2.0    & $<$0.80     \\
 33256 &6385&4.10&$-$0.30&3.12 & 1.32&  3.3:   & $<$3.0     &$<$1.40     \\
 33632A&5962&4.30&$-$0.23&4.40 & 1.06&  6.0  & 52.3   & 2.45   \\
 34411A&5773&4.02&   0.01&4.20 & 1.01&  9.4  & 27.4   & 2.02   \\
 35296A&6015&4.24&$-$0.14&4.15 & 1.00&  7.5  &106.4   & 2.87   \\
 38393A&6306&4.29&$-$0.07&3.83 & 1.21&  2.6  & 63.5   & 2.79   \\
 39587 &5805&4.29&$-$0.18&4.72 & 0.90& 10.7  &105.6   & 2.71   \\
 39833 &5767&4.06&   0.04&4.30 & 1.00&  9.4  & 61.7   & 2.40   \\
 41330 &5791&4.10&$-$0.24&4.01 & 0.98&  9.3  & 21.9   & 1.92   \\
 41640 &6004&4.37&$-$0.62&4.58 & 0.95&  7.2  & 54.7   & 2.51   \\
 43042 &6485&4.27&   0.04&3.58 & 1.30&  1.1  &  6.0   & 1.82   \\
 43947 &5859&4.23&$-$0.33&4.41 & 0.98&  9.8  & 47.0   & 2.33   \\
 46317 &6216&4.29&$-$0.24&3.80 & 1.19&  3.7  & 56.6   & 2.67   \\
 49732 &6260&4.15&$-$0.70&3.73 & 1.04&  6.3  & 29.7   & 2.37   \\
 51530 &6017&3.91&$-$0.56&2.82 & 1.28&  3.4     & $<$3.0    & $<$1.20     \\
 54717 &6350&4.26&$-$0.44&3.86 & 1.10&  4.6  & 24.2   & 2.34   \\
\hline
\end{tabular}
\end{table}

\begin{table}
\setlength{\tabcolsep}{0.10cm}
\indent
{\bf Table1.}(continued)~~~~~~~~~~~~~~~~~~~~~~~~~~~~~~~~~~~~~~~~~~~~~~~~~~~~~~~~\\[3mm]
\begin{tabular}{lrrrllrrr}
\noalign{\smallskip}
\hline
\noalign{\smallskip}
Star & $\teff$ & $\logg$ & \ffe & $\Mv$ & Mass & Age &E.W.& $\nlia$ \\
HD   & K     &       &      &     &$M_{\odot}$ & Gyr  &  $\mA$ & (Li)   \\
\noalign{\smallskip}
\hline
\noalign{\smallskip}
 55575 &5802&4.36&$-$0.36&4.44 & 1.00& 10.6  & 13.5   & 1.71   \\
 58551 &6149&4.22&$-$0.54&4.09 & 1.04&  5.9  & 44.0   & 2.49   \\
 58855 &6286&4.31&$-$0.31&3.87 & 1.16&  3.6  & 20.3   & 2.22   \\
 59380 &6280&4.27&$-$0.17&3.64 & 1.20&  4.2  & 18.1   & 2.16   \\
 59984A&5900&4.18&$-$0.71&3.55 & 0.99&  7.8  & 47.3   & 2.36   \\
 60319 &5867&4.24&$-$0.85&4.37 & 0.80& 17.0  & 21.2   & 1.96   \\
 61421 &6671&4.02&$-$0.02&2.65 & 1.50&  2.1:    &$<$2.0     & $<$1.40    \\
 62301 &5837&4.23&$-$0.67&4.07 & 0.90& 12.2  & 26.2   & 2.04   \\
 63077 &5831&4.19&$-$0.78&4.45 & 0.80& 13.6  &  6.1   & 1.37   \\
 63333 &6057&4.23&$-$0.39&3.92 & 1.03&  7.5  & 36.3   & 2.35   \\
 68146A&6227&4.16&$-$0.09&3.77 & 1.10&  4.9  & 38.6   & 2.49   \\
 68284 &5832&3.91&$-$0.56&3.41 & 1.09&  6.4  & 54.4   & 2.39   \\
 69897 &6243&4.28&$-$0.28&3.85 & 1.19&  3.3  & 62.7   & 2.73   \\
 70110 &5949&3.96&   0.07&3.12 & 1.40&       & 55.8   & 2.48   \\
 72945A&6202&4.18&   0.00&3.83 & 1.20&  2.8  &105.9   & 2.99   \\
 74011 &5693&4.05&$-$0.65&4.08 & 0.91& 12.6  & 18.6   & 1.79   \\
 75332 &6130&4.32&   0.00&3.96 & 1.15&  3.9  &106.4   & 2.94   \\
 76349 &6004&4.21&$-$0.49&3.88 & 1.02&  8.1  & 46.4   & 2.42   \\
 76932 &5873&4.12&$-$0.91&4.17 & 0.87& 12.5  & 26.0   & 2.06   \\
 78418A&5625&3.98&$-$0.26&3.48 & 1.19&  5.9  & 33.7   & 2.03   \\
 79028 &5874&4.06&$-$0.05&3.71 & 0.95&  7.4  & 76.5   & 2.59   \\
 80218 &6092&4.14&$-$0.28&3.65 & 1.18&  4.9    & $<$2.0     &$<$1.00     \\
 82328 &6302&3.91&$-$0.20&2.53 & 1.40&  2.8: &112.5   & 3.10   \\
 83220 &6470&4.06&$-$0.49&3.52 & 1.20&  5.1    & $<$3.0     &$<$1.50     \\
 86560 &5845&4.13&$-$0.48&4.06 & 0.98&  9.4  & 31.7   & 2.13   \\
 89125A&6038&4.25&$-$0.36&4.03 & 1.07&  6.5  & 40.2   & 2.38   \\
 90839A&6051&4.36&$-$0.18&4.30 & 1.03&  5.2  & 60.5   & 2.59   \\
 91347 &5808&4.35&$-$0.48&4.72 & 0.90& 10.8  & 24.0   & 1.98   \\
 91752 &6483&4.02&$-$0.23&2.95 & 1.31&  2.9:    & $<$2.0    & $<$1.30     \\
 91889A&6020&4.15&$-$0.24&3.76 & 1.15&  5.8  & 44.6   & 2.42   \\
 94280 &6063&4.10&   0.06&3.96 & 1.13&  4.8  & 52.1   & 2.53   \\
 95128 &5731&4.16&$-$0.12&4.31 & 0.92& 12.1  & 18.0   & 1.79   \\
 95241 &5808&3.91&$-$0.30&2.74 & 1.40&  3.3     &$<$1.0     & $<$0.50    \\
 97916 &6445&4.16&$-$0.94&3.64:& 1.04&  5.5    &$<$5.0&$<$1.50    \\
 99747 &6631&4.12&$-$0.54&3.24 & 1.26&  2.8     &$<$5.0     & $<$1.80    \\
100180A&5866&4.12&$-$0.11&4.44 & 0.95&  9.2  & 57.0   & 2.43   \\
100446 &5967&4.29&$-$0.48&4.14 & 0.98&  8.8  & 29.3   & 2.18   \\
100563 &6423&4.31&$-$0.02&3.65 & 1.27&  1.6  & 48.6   & 2.72   \\
101676 &6102&4.09&$-$0.47&3.64 & 1.09&  6.1  & 35.1   & 2.36   \\
102634 &6333&4.22&   0.24&3.48 & 1.35&  2.3  & 35.9   & 2.53   \\
102870 &6068&4.09&   0.13&3.41 & 1.32&       & 15.3   & 1.96   \\
103723 &6062&4.33&$-$0.75&4.65:& 0.96&  4.8  & 35.1   & 2.33   \\
103799 &6169&4.02&$-$0.45&3.29 & 1.20&  4.8  & 21.9   & 2.18   \\
105004 &5832&4.32&$-$0.78&5.56 & 0.90&       & 21.7   & 1.95   \\
106038 &5939&4.23&$-$1.33&4.98:& 0.70& 20.7  & 64.1   & 2.51   \\
106516A&6135&4.34&$-$0.71&4.35 & 0.90&  9.3     & $<$2.0    & $<$1.10     \\
107113 &6373&4.07&$-$0.54&3.34 & 1.18&  4.1     & $<$4.0    & $<$1.50     \\
108134 &5761&4.17&$-$0.44&4.24 & 0.98& 10.3  & 23.0   & 1.93   \\
108510 &5929&4.31&$-$0.06&4.44 & 1.05&  5.2  & 48.8   & 2.39   \\
109303 &5905&4.10&$-$0.61&3.55 & 1.00&  8.4  & 47.3   & 2.37   \\
109358 &5750&4.30&$-$0.19&4.66 & 0.89& 12.4  &  9.0   & 1.49   \\
110897 &5757&4.31&$-$0.59&4.76 & 0.85& 14.5  & 23.0   & 1.92   \\
113679 &5595&3.98&$-$0.71&3.75 & 0.95& 10.4  & 32.9   & 1.99   \\
114642A&6355&3.91&$-$0.17&2.47 & 1.41&  2.8: & 94.4   & 3.03   \\
114710A&5877&4.24&$-$0.05&4.45 & 0.94&  9.6  & 66.3   & 2.51   \\
\hline
\end{tabular}
\end{table}

\begin{table}
\setlength{\tabcolsep}{0.10cm}
\indent
{\bf Table1.}(continued)~~~~~~~~~~~~~~~~~~~~~~~~~~~~~~\\[3mm]
\begin{tabular}{lrrrllrrr}
\noalign{\smallskip}
\hline
\noalign{\smallskip}
Star & $\teff$ & $\logg$ & \ffe & $\Mv$ & Mass & Age &E.W.& $\nlia$ \\
HD   & K     &       &      &     &$M_{\odot}$ & Gyr  &  $\mA$ & (Li)   \\
\noalign{\smallskip}
\hline
\noalign{\smallskip}
114762 &5821&4.15&$-$0.74&4.26 & 0.88& 12.9  & 25.1   & 2.01   \\
114837 &6241&4.18&$-$0.28&3.63 & 1.20&  4.2  & 22.7   & 2.24   \\
115383A&5866&4.03&   0.00&3.92 & 1.09&  5.7  & 84.6   & 2.64   \\
118244 &6234&4.13&$-$0.55&3.71 & 1.07&  5.8  & 16.0   & 2.07   \\
120162 &5823&4.33&$-$0.73&4.75 & 0.84& 13.5  & 35.0   & 2.16   \\
120559 &5405&4.40&$-$0.88&6.22:& 0.80&         & $<$2.0     &$<$1.00     \\
121004 &5622&4.31&$-$0.71&4.72:& 0.80& 18.3     & $<$4.0    & $<$1.00     \\
121560 &6059&4.35&$-$0.38&4.24 & 1.00&  7.1  & 41.1   & 2.40   \\
124244 &5853&4.11&   0.05&3.57 & 1.19&  5.8  & 51.9   & 2.37   \\
126512 &5714&4.02&$-$0.63&3.92 & 0.97& 10.1  & 18.6   & 1.81   \\
126681 &5533&4.28&$-$1.14&5.59:& 0.70& 17.5  & 14.9   & 1.54   \\
127334 &5501&4.16&   0.05&4.52 & 0.90& 15.9  & 16.4   & 1.60   \\
128167 &6737&4.25&$-$0.41&3.53 & 1.30&  1.0:   &$<$3.0&$<$1.60    \\
128385 &6041&4.12&$-$0.33&4.26 & 1.06&  5.5  & 31.0   & 2.26   \\
130551 &6274&4.19&$-$0.72&3.77 & 1.07&  7.6  & 28.2   & 2.36   \\
130948 &5780&4.18&$-$0.20&4.61 & 0.97& 10.0  &103.1   & 2.69   \\
131117 &6014&4.00&   0.13&3.17 & 1.40&       & 73.2   & 2.67   \\
132254 &6231&4.22&   0.07&3.64 & 1.23&  3.2  & 35.4   & 2.45   \\
136351 &6343&3.91&   0.01&2.32 & 1.62&       & 37.6   & 2.56   \\
139457 &5941&4.06&$-$0.52&3.70 & 0.98&  8.7  & 48.8   & 2.41   \\
141004 &5801&4.16&$-$0.04&4.07 & 1.04&  8.4  & 23.0   & 1.95   \\
142373 &5920&4.27&$-$0.39&3.61 & 0.96&  9.7  & 61.5   & 2.51   \\
142860A&6227&4.18&$-$0.22&3.63 & 1.20&  4.2  & 19.0   & 2.15   \\
143761 &5698&4.14&$-$0.26&4.20 & 0.99& 11.2  &  6.0   & 1.28   \\
146099A&5941&4.10&$-$0.61&3.57 & 1.00&  8.1  & 45.5   & 2.37   \\
146588 &5895&4.29&$-$0.63&4.53 & 0.90& 11.0  & 42.6   & 2.31   \\
149750 &5792&4.17&   0.08&4.05 & 1.11&  7.5  & 30.6   & 2.07   \\
150453 &6458&3.91&$-$0.37&2.41 & 1.51&  2.3: & 32.8   & 2.56   \\
154417 &5925&4.30&$-$0.04&4.47 & 1.05&  5.0  & 80.5   & 2.65   \\
155358 &5818&4.09&$-$0.67&4.09 & 0.89& 12.7  & 23.0   & 1.97   \\
157347 &5654&4.36&$-$0.02&4.84 & 1.06&  7.7     & $<$5.0    & $<$1.10     \\
157466 &5935&4.32&$-$0.44&4.51 & 0.95&  8.4  & 46.7   & 2.38   \\
159332 &6204&3.91&$-$0.23&2.82 & 1.29&  3.3:    & $<$4.0    & $<$1.50     \\
160291 &6008&4.10&$-$0.53&3.78 & 1.04&  7.5  & 28.4   & 2.20   \\
162004B&6059&4.12&$-$0.08&4.03 & 1.10&  5.6  & 65.9   & 2.64   \\
165908 &6083&4.19&$-$0.56&3.98 & 1.04&  6.8  & 38.2   & 2.39   \\
167588 &5894&4.13&$-$0.33&3.45 & 1.18&  4.9  & 46.5   & 2.35   \\
168009 &5719&4.08&$-$0.07&4.52 & 0.89& 12.8    & $<$4.0     &$<$1.10     \\
168151 &6529&4.15&$-$0.31&3.17 & 1.35&  2.5     &  $<$2.0   & $<$1.40      \\
170153A&6034&4.28&$-$0.65&4.04 & 0.95&  9.4  & 36.0   & 2.32   \\
174912 &5787&4.35&$-$0.54&4.77 & 0.87& 13.8  & 32.8   & 2.11   \\
184601 &5830&4.20&$-$0.81&4.48:& 0.80& 17.1  & 40.7   & 2.24   \\
186379 &5816&3.99&$-$0.46&3.60 & 1.10&  7.0  & 44.8   & 2.29   \\
187691 &6034&4.16&   0.09&3.69 & 1.23&  4.5: & 69.9   & 2.65   \\
189340 &5888&4.26&$-$0.19&3.93 & 0.96&  9.4  & 46.0   & 2.34   \\
191862A&6328&4.19&$-$0.27&3.61 & 1.15&  5.9  & 62.9   & 2.79   \\
198390 &6339&4.20&$-$0.31&3.62 & 1.20&  3.8  & 17.0   & 2.17   \\
199960 &5750&4.17&   0.11&4.10 & 1.09&  8.3  & 59.0   & 2.37   \\
200580 &5829&4.39&$-$0.58&3.58 & 0.87& 13.3  & 22.9   & 1.97   \\
201891 &5827&4.43&$-$1.04&4.64 & 0.70& 21.7  & 25.5   & 2.01   \\
203608 &6109&4.34&$-$0.67&4.41 & 0.95&  7.2  & 37.3   & 2.39   \\
204306 &5896&4.09&$-$0.65&3.74 & 0.98&  8.9  & 44.1   & 2.33   \\
204363 &6141&4.18&$-$0.49&3.89 & 1.06&  6.3  & 73.3   & 2.75   \\
206301 &5682&3.98&$-$0.04&3.68:& 1.09&  6.7  & 62.1   & 2.35   \\
206860 &5798&4.25&$-$0.20&4.66 & 0.97&  9.9  &109.5   & 2.73   \\
208906A&5929&4.39&$-$0.73&4.62 & 0.83& 13.6  & 40.7   & 2.31   \\
\hline
\end{tabular}
\end{table}

\begin{table}
\setlength{\tabcolsep}{0.10cm}
\indent
{\bf Table1.}(continued)~~~~~~~~~~~~~~~~~~~~~~~~~~\\[3mm]
\begin{tabular}{lrrrllrrr}
\noalign{\smallskip}
\hline
\noalign{\smallskip}
Star & $\teff$ & $\logg$ & \ffe & $\Mv$ & Mass & Age &E.W.& $\nlia$ \\
HD   & K     &       &      &     &$M_{\odot}$ & Gyr  &  $\mA$ & (Li)   \\
\noalign{\smallskip}
\hline
\noalign{\smallskip}
209942A&6022&4.25&$-$0.29&4.00 & 1.10&  6.2  & 38.5   & 2.35   \\
210027A&6496&4.25&$-$0.17&3.41 & 1.23&  2.7  & 72.0   & 2.98   \\
210752 &5847&4.33&$-$0.68&4.56 & 0.85& 14.0  & 39.0   & 2.23   \\
212029A&5875&4.36&$-$1.01&4.70 & 0.76& 18.8  & 27.9   & 2.08   \\
215257 &5976&4.36&$-$0.65&4.28 & 0.90& 11.2  & 38.3   & 2.31   \\
216385 &6244&3.97&$-$0.25&3.02 & 1.29&  3.3:    & $<$2.0    & $<$1.20     \\
218470 &6495&4.06&$-$0.13&3.04 & 1.32&  2.6:     &  $<$4.0  &  $<$1.70      \\
219476 &5887&3.91&$-$0.59&2.78 & 1.27&  3.4      &  $<$2.0  &  $<$0.90      \\
219623A&6039&4.07&   0.02&4.04 & 1.10&  5.5  & 68.0   & 2.65   \\
221377 &6348&3.91&$-$0.96&2.80 & 1.13&  4.5      &  $<$3.0  &  $<$1.40      \\
222368 &6169&4.06&$-$0.17&3.43 & 1.15&  4.8  & 20.8   & 2.16   \\
241253 &5830&4.23&$-$1.10&5.36:& 0.76& 10.8  & 29.5   & 2.08   \\
BD-21\,3420&5858&4.25&$-$1.09&5.08:& 0.72& 19.2  & 21.5   & 1.95   \\
CD-33\,3337 &6022&3.99&$-$1.34&4.16:& 0.77& 16.4  & 38.0   & 2.32   \\
CD-45\,3283 &5650&4.50&$-$0.84&5.71:& 0.80&         &  $<$2.0   & $<$0.70    \\
CD-47\,1087 &5657&4.20&$-$0.80&5.08:& 0.74& 22.7  &  5.7   & 1.21   \\
CD-57\,1633 &5944&4.22&$-$0.89&5.12:& 0.90&       & 33.7   & 2.22   \\
CD-61\,0282 &5772&4.20&$-$1.23&5.27:& 0.70& 18.6  & 34.8   & 2.12   \\
G005-040 &5737&4.02&$-$0.91&4.76:& 0.77& 19.5  & 10.1   & 1.90   \\
G046-031 &5907&4.18&$-$0.81&5.12:& 0.88&      & 26.8   & 2.09   \\
G088-040 &5911&4.14&$-$0.83&4.60:& 0.83& 13.8  & 24.6   & 2.06   \\
G102-020 &5310&4.56&$-$1.09&6.00:& 0.70&       &  7.1   & 1.26   \\
  W7547 &6272&4.03&$-$0.42&3.27:& 1.25&         & $<$2.0    & $<$1.20     \\
\hline
\noalign{\smallskip}
\end{tabular}
\end{table}

\end{document}